\documentclass{article}
\usepackage{ijcai25}

\usepackage{times}
\usepackage{soul}
\usepackage{url}
\usepackage[hidelinks]{hyperref}
\usepackage[utf8]{inputenc}
\usepackage[small]{caption}
\usepackage{graphicx}
\usepackage{amsmath}
\usepackage{amsthm}
\usepackage{booktabs}
\usepackage{algorithm}
\usepackage{algorithmic}
\usepackage[switch]{lineno}

\usepackage{multirow}
\usepackage{amssymb}
\usepackage{graphicx}
\usepackage{subfig}

\title{RTLRepoCoder: Repository-Level RTL Code Completion through the Combination of Fine-Tuning and Retrieval Augmentation}

\author{
    Peiyang Wu,
    Nan Guo,
    Junliang Lv,
    Xiao Xiao,
    Xiaochun Ye
    \affiliations
    Institute of Computing Technology, Chinese Academy of Sciences
    \emails
    \{wupeiyang,guonan\}@ict.ac.cn
}

\begin{document}

\maketitle

\begin{abstract}
    As an essential part of modern hardware design, manually writing Register Transfer Level (RTL) code such as Verilog is often labor-intensive. Following the tremendous success of large language models (LLMs), researchers have begun to explore utilizing LLMs for generating RTL code. However, current studies primarily focus on generating simple single modules, which can not meet the demands in real world. In fact, due to challenges  in managing  long-context RTL code and complex cross-file dependencies,  existing solutions cannot handle large-scale Verilog repositories in practical hardware development. As the first endeavor to exclusively adapt  LLMs for large-scale RTL development, we propose RTLRepoCoder, a groundbreaking solution that incorporates specific fine-tuning and Retrieval-Augmented Generation (RAG) for repository-level Verilog code completion. Open-source Verilog repositories from the real world, along with an extended context size, are used for domain-specific fine-tuning. The optimized RAG system improves the information density of the input context by retrieving relevant code snippets. Tailored optimizations for RAG are carried out, including the embedding model, the cross-file context splitting strategy, and the chunk size. Our solution achieves state-of-the-art performance on public benchmark, significantly surpassing GPT-4 and advanced domain-specific LLMs on Edit Similarity and Exact Match rate. Comprehensive experiments demonstrate the  remarkable  effectiveness of our approach and offer insights for future work. 
\end{abstract}

\section{Introduction}
Modern digital design typically relies on writing Register Transfer Level (RTL) code like Verilog, which bridges high-level functional specifications with low-level circuit implementations. However, manually writing RTL code is time-intensive and shifts developers' focus from higher-level system design to low-level implementation details. With the remarkable success of large language models (LLMs) in high-level language generation like Python\cite{guo2024deepseek,li2023starcoder,roziere2023code}, their application to RTL code has emerged as a promising direction.

A popular approach is to fine-tune foundation LLMs using data related to RTL.  As an example, VeriGen\cite{thakur2024verigen} finetunes 16B CodeGen\cite{nijkamp2022codegen} model utilizing data from GitHub and Verilog textbooks. RTLCoder\cite{liu2024rtlcoder} synthesizes description-code pairs by prompting GPT-3.5 and utilizes them for instruction tuning. Although existing approaches can correctly generate simple modules, they struggle to handle complex requirements from real world.

In most real-world development processes, developers have to face complex code repositories rather than single file or single module. These repositories are typically lengthy and involve intricate cross-module dependencies. Current RTL LLMs are limited by relatively short context sizes (e.g. 4k in RTLCoder\cite{liu2024rtlcoder}) during fine-tuning, making it challenging to process long contexts. Furthermore, the simple fine-tuning data they use limits models' ability to capture cross-module dependencies effectively.

To address the aforementioned issues, we propose RTLRepoCoder, a novel solution for repository-level RTL code completion. Specifically, to tackle the limitations of existing RTL LLMs in handling long contexts and cross-module dependencies, we fine-tune the LLM on Verilog repositories from GitHub with a context length of 10k. By expanding the context window size, the fine-tuned model significantly enhances the ability for repository-level RTL code completion. Furthermore, for contexts exceeding the preset context length, we adopt the Retrieval-Augmented Generation (RAG) framework. Relying on an embedding model, relevant code snippets are retrieved from cross-file contexts and concatenated with in-file contexts before feeding them into the LLM. We carefully analyze the selection of the embedding model, cross-file contexts splitting strategies and chunk sizes to make specific optimizations for Verilog repositories. By fusing fine-tuning and RAG organically, we propose RTLRepoCoder, a novel solution for repository-level RTL code completion, which holds great potential for application in the complex real-world development scenarios. The contributions of RTLRepoCoder can be summarized as follows:
\begin{itemize}
\item To the best of our knowledge, we propose the first solution specifically designed for repository-level RTL code completion, combining domain-specific fine-tuning and optimized RAG. This comprehensive solution holds great potential for facilitating the hardware design in the real world.
\item We fine-tune an LLM using RTL repositories from Github, empowering it to process  long Verilog code, handle cross-file, cross-module dependencies and predict the next line of code. 
\item We introduce a RAG system and perform detailed optimizations, significantly enhancing the model's ability to complete RTL code that exceeds the length limit.
\item Comprehensive experiments demonstrate the superiority of our approach and the notable effectiveness of each component. Our solution achieves SOTA performance on the RTL-Repo public benchmark\cite{allam2024rtl},  with a substantial margin over GPT-4 and other advanced domain-specific LLMs.
\end{itemize}

The rest of this article is structured as follows. Section 2 briefly reviews related works on LLMs for RTL code generation and repository-level code completion. Section 3 presents our solution, including the overall workflow, fine-tuning process and the RAG system. Section 4 introduces the experimental setup. And Section 5 provides comprehensive experiment results, and offers in-depth analyses. Finally, Section 6 concludes the paper and discusses future directions.

\section{Related Works}

\subsection{LLMs for RTL Code Generation}

Some researchers try to directly prompt off-the-shelf LLMs  to generate RTL code. Chip-Chat\cite{blocklove2023chip} and ChipGPT\cite{chang2023chipgpt} leverage commercial LLMs to generate RTL code with the aid of experienced engineers, thus limiting the degree of automation. AutoChip\cite{thakur2023autochip} employs error reports from compilers and simulators to help LLMs correct faulty code. 

To enhance the expertise of LLMs in RTL code generation, fine-tuning with domain-specific data has become a mainstream approach. As a pioneer, VeriGen\cite{thakur2024verigen} fine-tune the CodeGen-16B\cite{nijkamp2022codegen} model using Verilog data from GitHub and textbooks. ChipNeMo\cite{liu2023chipnemo} tailors LLMs for applications including chatbot, generating EDA tool script, and bug summarization. RTLCoder\cite{liu2024rtlcoder} proposes an automated flow to generate a large instruction fine-tuning dataset and introduces a novel LLM training approach that incorporates code quality feedback. CodeV\cite{zhao2024codev} propose an approach to construct an instruction-tuning dataset by using GPT-3.5 to summarize Verilog code into multi-level descriptions. MEV-LLM\cite{nadimi2024multi} presents  a multi-expert architecture to integrate multiple LLMs fine-tuned on datasets with varying design complexities. BetterV\cite{pei2024betterv} fine-tunes LLMs on specialized datasets and employs generative discriminators for design guidance, resulting in improved functional correctness and enhanced performance in EDA downstream tasks. Although the above studies have made some progress, they mainly focus on simple single-module, single-file RTL generation, which falls short of meeting the needs of large-scale repository-level development in real-world scenarios.

\subsection{Repository-level Code Completion}
Repository-level code completion for high-level languages has gained significant attention, as it tackles the real-world challenge of generating code by considering cross-file context. Repo-Level Prompt Generator\cite{shrivastava2023repository}  creates example-specific prompts leveraging proposals that capture the repository's structure and context from relevant files. RepoCoder\cite{zhang2023repocoder} iteratively retrieves relevant code snippets from repositories and completes the code, significantly improving performance. Furthermore, RepoHyper\cite{phan2024repohyper} introduces a novel semantic graph representation to improve context retrieval accuracy. Unlike fully relying on RAG, RepoFusion\cite{shrivastava2023repofusion} fine-tunes LLMs using  relevant contexts extracted from repositories. CrossCodeEval\cite{ding2024crosscodeeval} and RepoBench\cite{liu2023repobench} serve as public benchmarks for evaluating repository-level code completion performance.

In contrast, repository-level code completion for RTL is still under research. To the best of our knowledge, there is currently no dedicated solution or model proposed to address repository-level code completion for hardware description languages like Verilog. RTL-Repo\cite{allam2024rtl}, as the only existing contribution in this area, presents a benchmark that contains 4,098 samples from open-source Verilog repositories on GitHub, with the context length ranging from 2k to 128k.

\section{Method}
In this section, we first introduce the task formulation and the overall pipeline, and then explain the fine-tuning and RAG system in detail.

\subsection{Task Formulation and Overall Framework}

\begin{figure}[htbp]
  \centering
  \includegraphics[width=\linewidth]{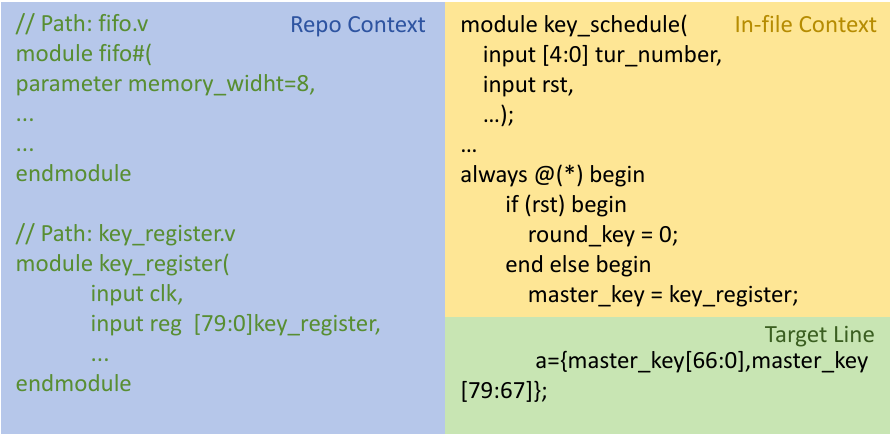}
  \caption{An illustrative example of repository-level RTL code completion. Due to space limitations, only a portion of the Verilog repository is shown.}
  \label{fig:visual_code}
  
\end{figure}

\begin{figure*}[!htbp]
  \centering
  \includegraphics[width=.9\linewidth]{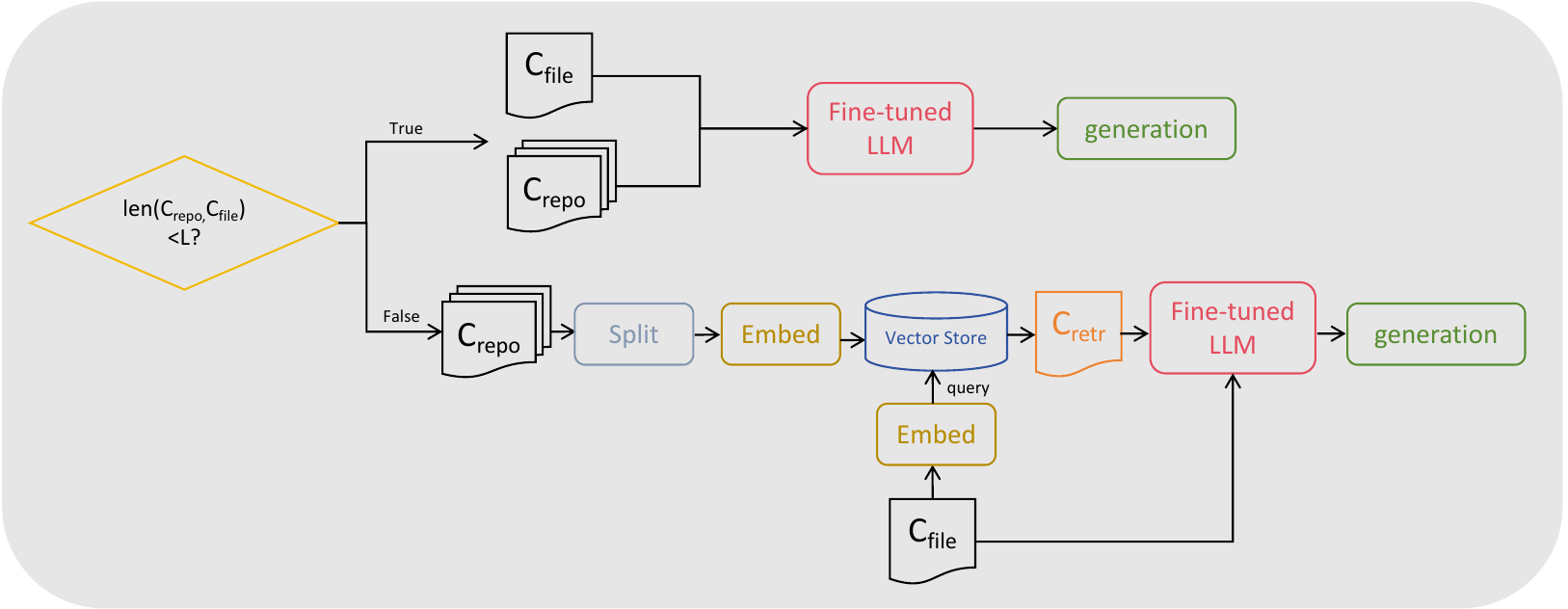}
  \caption{The overall pipeline of RTLRepoCoder}
  \label{fig:pipeline}
  
\end{figure*}

As Figure~\ref{fig:visual_code} shows, in the repository-level RTL code completion task, the model needs to predict the next line of code based on the context within the repository $C_{repo} $ and the incomplete content of the current file $C_{file} $. Following previous work\cite{allam2024rtl}, the model's performance is evaluated based on its ability to predict the next line of code. The prediction process can be formalized as: 
\begin{equation}\label{eqn:setup}
Y=\mathcal{M}(C_{repo},C_{file}  ) 
\end{equation}

This evaluation  approach is effective because the model need to understand the repository's overall structure, logic, and coding style to accurately generate the next line. 

  

Figure~\ref{fig:pipeline} illustrates our overall framework. Firstly, we determine the generation strategy based on the length of the input context. If the token length of the concatenation of $C_{repo}$ and $C_{file}$ is below the preset limit $L$ (e.g. context length used for fine-tuning), they are directly fed into the model. Otherwise, we apply the RAG strategy: $C_{file}$ is used to retrieve relevant code snippets $C_{retr}$ from $C_{repo}$. The retrieval process can be formalized as $C_{retr}=\mathcal{R}(C_{repo},C_{file}  ) $. Then the concatenation of $C_{retr}$ and $C_{file}$ is passed into the model. The overall process can be formalized as follows:

\begin{equation}\label{eqn:overall}
Y= 
\left\{ 
    \begin{array}{lc}
        \mathcal{M}(C_{repo},C_{file}  ) &  len(Cat(C_{repo},C_{file})) < L \\
        \mathcal{M}(C_{retr},C_{file}  )  &else\\
    \end{array}
\right.
\end{equation}

To enhance the performance of the entire process, we fine-tune the LLM $\mathcal{M}$ and optimize the retrieval process $\mathcal{R}$. In fact, these two approaches complement each other: fine-tuning effectively enhances the model's ability to process code within the predefined context size, while RAG  increases the information density within the context size by retrieving relevant code snippets.

\subsection{Fine-tuning Process}

\begin{table*}[htbp]
\caption{Main results on the RTL-Repo\protect\cite{allam2024rtl} benchmark. The results of GPT-3.5, GPT-4, Starcoder2\protect\cite{lozhkov2024starcoder}, VeriGen\protect\cite{thakur2024verigen}, RTLCoder-Mistral\protect\cite{liu2024rtlcoder} and RTLCoder-DeepSeek\protect\cite{liu2024rtlcoder} are referenced from the original paper\protect\cite{allam2024rtl}. According to the article, GPT-4 was evaluated on 200 randomly selected samples due to budget constraints. The results are presented as percentages(\%), with the top performance emphasized in bold and the second-best result underlined.}
\centering
\begin{tabular}{@{}lcccc@{}}
\hline
\textbf{Model}               & \textbf{Type}           & \textbf{Num of Parameters} & \textbf{Edit Similarity} & \textbf{Exact Match} \\ \hline
GPT-3.5                      & General-Purpose         & N/A                        & 61.4                     & 33.8                \\
GPT-4                    & General-Purpose         & N/A                        & \underline{71.9}           & \underline{48.5}       \\
Starcoder2\cite{lozhkov2024starcoder}               & General-Purpose Code         & 15B                        & 48.0                     & 17.0                \\ \hline
VeriGen\cite{thakur2024verigen}                  & Verilog-Specific        & 16B                        & 43.9                     & 9.5                 \\
RTLCoder-Mistral\cite{liu2024rtlcoder}         & Verilog-Specific        & 7B                         & 37.9                     & 8.3                 \\
RTLCoder-DeepSeek\cite{liu2024rtlcoder}        & Verilog-Specific        & 6.7B                       & 48.1                     & 16.2                \\
RTLRepoCoder-DeepSeek        & Verilog-Specific        & 6.7B                       & \textbf{84.3}                     & \textbf{55.8}                \\ \hline
\end{tabular}

\label{tab:model_comparison}
\end{table*}

The instruction fine-tuning is carried out to enable the model to understand long-contexts and cross-file dependencies, and accurately predict the next line of code. Specifically, $C_{repo}$ and $C_{file}$ are concatenated to form the inputs, with the target next line $\hat{Y} $ being the label. The LLM is optimized based on the cross-entropy loss between the predicted probabilities and the target labels:

\begin{equation}\label{eqn:ft}
Loss=-\frac{1}{\left \| \hat{Y}  \right \| } \sum_{t} \log p_\theta \left( \hat{Y}_t \mid C_{repo},C_{file}, \hat{Y}_{<t} \right),
\end{equation}

where t represents the index of the token in $\hat{Y}$, and $\left \| \hat{Y}  \right \|$ represents the length of $\hat{Y}$. To adapt LLMs to the complex situations in real-world scenarios, we utilize data from open-source Verilog repositories on GitHub. Additionally, to enable the LLM to handle long contexts in repositories, we employ a 10,240 context size during fine-tuning, which significantly surpasses previous works\cite{thakur2024verigen,liu2024rtlcoder} with context window sizes of 4,096 or even smaller.

\subsection{RAG system}

For inputs exceeding the predefined length during the inference stage, we adopt a RAG approach to retrieve relevant code snippets, thereby preventing input length overflow. As Figure~\ref{fig:pipeline} shows, the context within the repository $C_{repo}$ is first split into chunks, which are subsequently encoded by an embedding model and stored in a vector store. Subsequently, the embedding of $C_{file}$ is used as the query to retrieve relevant chunks from the vector store based on cosine similarity. The retrieved chunks are then added to $C_{retr}$ in descending order of relevance until the input length reaches the predefined length $L$.

A series of optimizations are carried out to improve the RAG system. First, considering the characteristics of Verilog language, we choose to split chunks based on the keyword '\textbackslash n' (line break) to minimize the disruption to the  intrinsic structure of repositories. Second, since the length of meaningful Verilog code snippet might exceed the maximum length support by typical embedding models like bge-large-en-v1.5\cite{xiao2024c}, we introduce the jina-embeddings-v2-base-en model\cite{gunther2023jina}, which supports 8k context length. Third, we investigate the impact of chunk size on the quality of the generated results.

\section{Experimental Setup}

In this section, we introduce the experimental setup, including the dataset, evaluation metrics, and the implementation details of the entire solution.

\subsection{Dataset}
We conduct training and performance evaluation on the RTL-Repo\cite{allam2024rtl} dataset, which contains 4,098 samples collected from 1,361 GitHub Verilog repositories. Each sample is organized into three parts: code from the repository $C_{repo} $, incomplete code from current file $C_{file} $, and the target next line $\hat{Y} $. The token length of the context varies from 2k to 128k. The dataset is divided into a training set with 2,924 samples and a test set with 1,174 samples. We randomly select 2,000 samples from the training set for training and evaluate performance on the full test set.

\subsection{Evaluation Metrics}

Following related work\cite{lu2021codexglue} in high-level programming languages, Exact Match (EM) and Edit Similarity (ES) are used as evaluation metrics. EM measures the proportion of predictions $Y $ that are exactly identical to the human-written target line $\hat{Y} $. Considering the flexibility of coding, ES serves as a smoother metric, measuring the similarity between the prediction $Y $ and the human-written target line $\hat{Y} $ based on the Levenshtein distance\cite{levenshtein1966binary}.

\subsection{Implementation Details}
We adopt a context window size of 10,240 tokens in both the fine-tuning and inference processes. The sampling temperature is set to 0.2. DeepSeek-Coder-Instruct-6.7B \cite{guo2024deepseek} is selected as base models. The experiments are conducted on 8 NVIDIA A800 80GB GPUs. The LLM is fine-tuned for 1 epoch with a batch size of 2 per GPU and gradient accumulation over 2 steps. Stage 3 of the ZeRO\cite{rajbhandari2020zero} mechanism is introduced during fine-tuning to prevent GPU memory overflow caused by long contexts. 

For the RAG process, we implement splitting and semantic retrieval using LangChain. The chunk size is set to 4,096 tokens. Jina-embeddings-v2-base-en model\cite{gunther2023jina} is selected as the embedding model for its  capability to support a long context length of 8,192 tokens. The vector store is built with Milvus\cite{wang2021milvus}, an popular high-performance vector database.

\section{Results and Analysis}

In this section, detailed experimental results and in-depth analyses are provided, including main results, discussion of fine-tuning and RAG, and the ablation study of the initial decision-making process.

\subsection{Main Results}

Table~\ref{tab:model_comparison} shows the main results on RTL-Repo\cite{allam2024rtl} benchmark. The results of various LLMs are presented, including general-purpose LLMs, general-purpose code LLM, and Verilog-specific LLMs. Compared to other LLMs, our model achieves SOTA performance on both EM and ES metrics, even surpassing GPT-4. Compared to RTLCoder-DeepSeek\cite{liu2024rtlcoder}, which is built on the same base model as us, our solution demonstrates remarkable superiority, achieving absolute gains of 36.2\% in ES  and 39.6\% in EM. Moreover, our model is highly parameter-efficient, delivering remarkable performance with only 6.7B parameters, outperforming larger models such as StarCoder2 (15B) \cite{lozhkov2024starcoder} and VeriGen (16B) \cite{thakur2024verigen}. 
Overall, as the first solution designed for repository-level RTL code completion, our proposed approach achieves SOTA performance while maintaining exceptional parameter efficiency.

\subsection{Discussion of Fine-tuning}

\subsubsection{Impact of Fine-tuning}\label{sec:ft}

\begin{figure}[htbp]
  \centering
  \includegraphics[width=\linewidth]{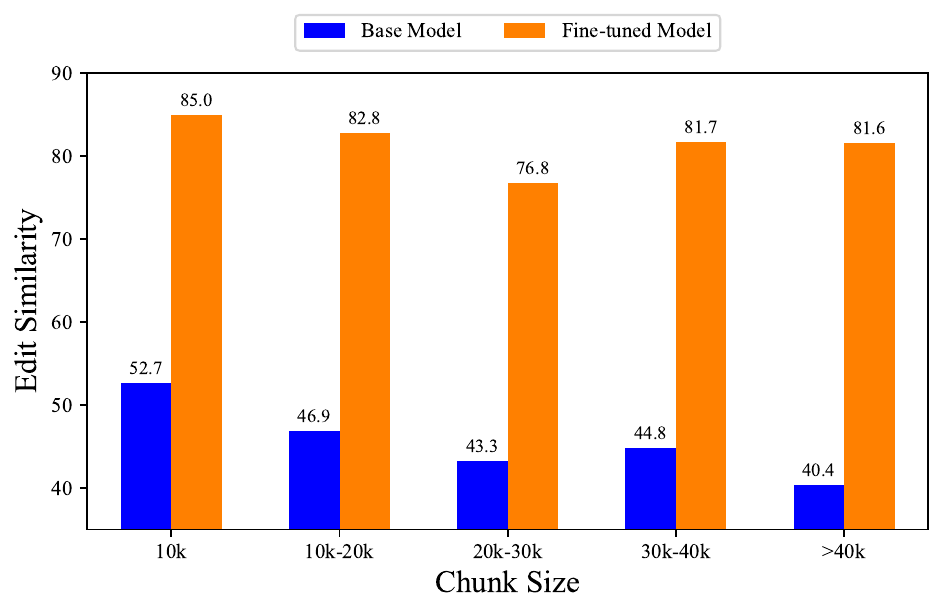}
  \caption{Edit Similarity of LLMs before and after fine-tuning across samples with different context lengths. The results are presented as percentages(\%).}
  \label{fig:ft}
  
\end{figure}

To investigate the improvements brought by fine-tuning, we compare the model's performance before and after fine-tuning. To ensure a fair comparison, the experiment is conducted without utilizing any retrieval augmentation. Based on the performance shown in Table~\ref{tab:ft}, fine-tuning  with Verilog repositories results in absolute increases of 34.7\% in ES and 37.1\% in EM. Specifically, as presented in Figure~\ref{fig:ft}, fine-tuning leads to substantial improvements across samples with different context lengths. Through fine-tuning, the model enhances its ability to understand long code within the repository and predict the next line. However, it is also evident that, due to the context length limitation, the model's completion ability for samples longer than 10k tokens is noticeably weaker compared to those shorter than 10k. This issue can be mitigated through the use of RAG. 

\begin{table}[ht]
    \centering
    \begin{tabular}{ccccc}
        \hline
        \textbf{Fine-tuning with Verilog Repositories} & \textbf{ES} & \textbf{EM} \\
        \hline
        $\times$ & 48.6 & 16.8 \\

        \checkmark & \textbf{83.3} & \textbf{53.9} \\
        \hline
    \end{tabular}
    \caption{Performance before and after fine-tuning with Verilog repositories. The results are presented as percentages(\%), with the top performance emphasized in bold.}
    \label{tab:ft}
\end{table}

\begin{figure}[h]
  \centering
  \begin{minipage}[t]{.48\linewidth}
    \centering
    \subfloat[]{\label{incorrect, non-self-contained code}\includegraphics[width=\textwidth]{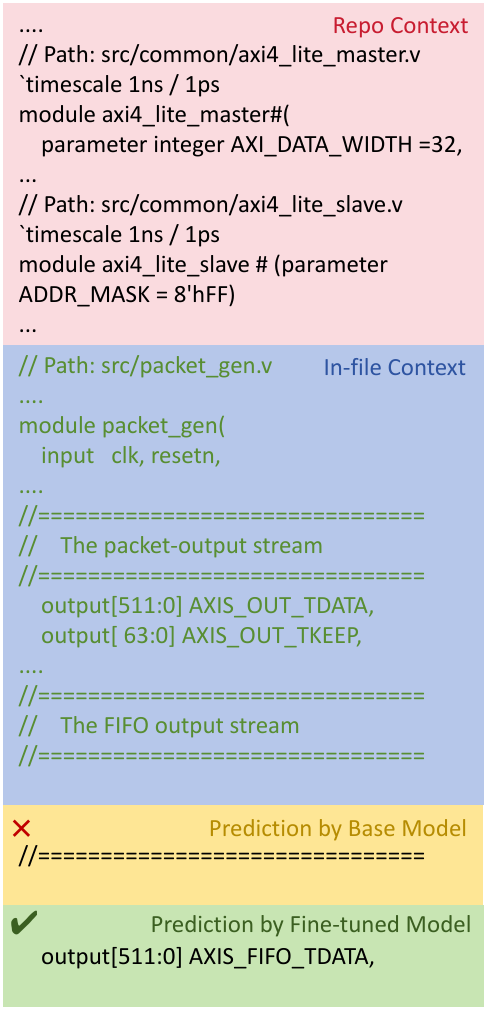}}

\end{minipage} 
\begin{minipage}[t]{.48\linewidth}
    \centering
    \subfloat[]{\label{incorrect, non-self-contained code}\includegraphics[width=\textwidth]{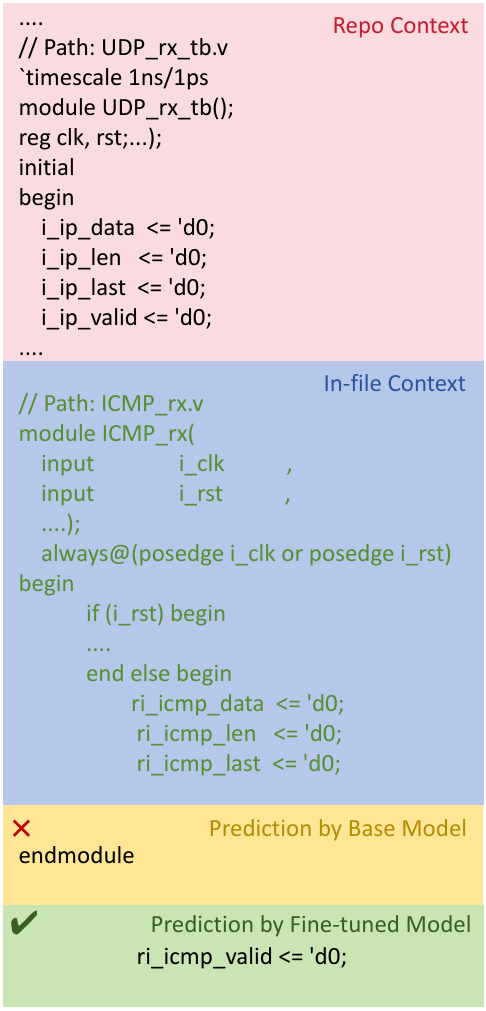}}

\end{minipage} 
\caption{Two examples showing the prediction changes before and after fine-tuning. The base model outputs irrelevant comment information (a) or prematurely terminate the module (b), while the fine-tuned model generates correct Verilog code.}
\label{fig:ft_code}
  
\end{figure}

To gain a deeper understanding of the impact of fine-tuning, we carefully compare the LLM's predictions before and after fine-tuning.  A key finding is that, before fine-tuning, the base model often generates irrelevant content, such as adding comments or prematurely terminating modules, indicating that it has not fully adapted to the specific task. This issue is considerably improved after fine-tuning.  Figure~\ref{fig:ft_code} shows two examples\cite{indy}. Before fine-tuning, the base model outputs a meaningless comment (a) or 'endmodule' (b).  After fine-tuning, the model successfully generates the correct port definition (a) and assignment expression (b), demonstrating that fine-tuning standardizes the LLM's output.

  

\subsubsection{Context Size}

Compared to previous work\cite{liu2024rtlcoder,thakur2024verigen} that used context window sizes of 4,096 or even smaller during fine-tuning, we choose to expand the window size to 10,240 to enhance the LLM's ability to handle repository-level long Verilog code. To validate the effectiveness of this idea, we conduct an experiment by adjusting the context window size during fine-tuning and fixing it during inference to ensure fairness in the comparison. As the results presented in Table~\ref{tab:con_size} indicate, both ES and EM metrics significantly improve with an increase in the context window size from 4,096 to 10,240 during fine-tuning. This suggests that increasing the context window size during fine-tuning helps the LLM better adapt to repository-level  long Verilog  code.

\begin{table}[ht]
    \centering
    \begin{tabular}{ccccc}
        \hline
        \textbf{Context Size of Fine-tuning} & \textbf{ES} & \textbf{EM} \\
        \hline
        4096 & 82.0 & 52.9 \\

        
        10240 & \textbf{83.3} & \textbf{53.9} \\
        \hline
    \end{tabular}
    \caption{Performance  with different context size during fine-tuning.  The context window size during inference is fixed at 10,240 to ensure fairness in the comparison. The results are presented as percentages(\%), with the top performance emphasized in bold.}
    \label{tab:con_size}
\end{table}

\subsection{Discussion of RAG}

\subsubsection{Ablation Study of RAG}
\begin{figure}[htbp]
  \centering
  \includegraphics[width=\linewidth]{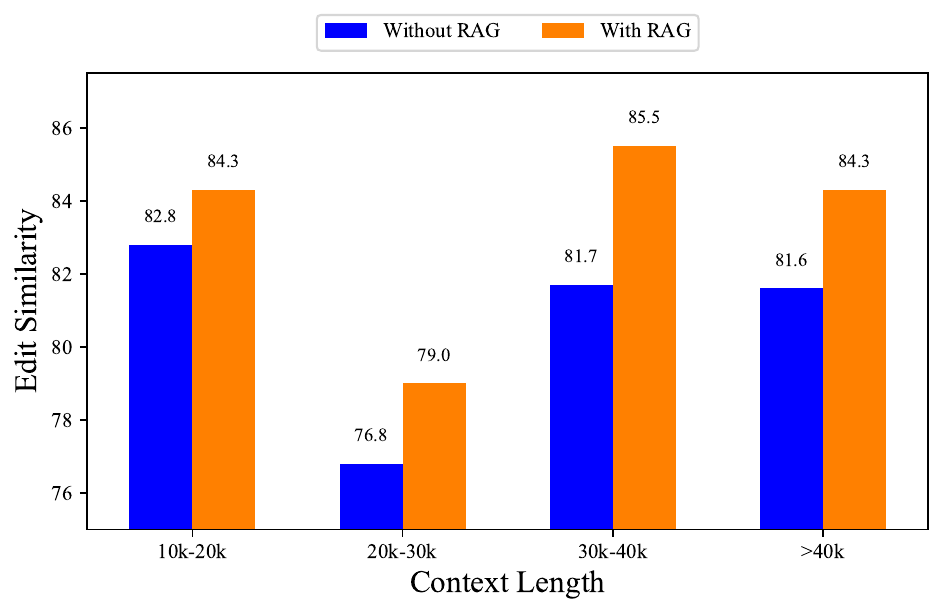}
  \caption{Edit Similarity of LLMs with and without RAG on samples exceeding 10k tokens of context. The results are presented as percentages(\%).}
  \label{fig:rag}
  
\end{figure}

To evaluate the effectiveness of RAG, we conduct an ablation study on samples  whose token length exceeds 10k, where the RAG mechanism is triggered. 
According to the results in Table~\ref{tab:rag}, the introduction of RAG significantly enhances the model's predictive capability on samples that exceed the predefined length. Furthermore, as shown in Figure~\ref{fig:rag}, as the context length increases, there is no obvious downward trend in performance. This phenomenon demonstrates that the RAG mechanism effectively retrieves relevant code snippets, providing the LLM with valuable information for RTL code completion.

\begin{table}[ht]
    \centering
    \begin{tabular}{ccccc}
        \hline
        \textbf{Retrival Augmentation} & \textbf{ES} & \textbf{EM} \\
        \hline
        $\times$ & 81.2 & 50.3 \\

        \checkmark & \textbf{83.5} & \textbf{55.0}  \\
        \hline
    \end{tabular}
    \caption{Performance on samples exceeding 10k tokens of context with and without retrival augmentation. The results are presented as percentages(\%), with the top performance emphasized in bold.}
    \label{tab:rag}
\end{table}

\begin{figure}[h]
  \centering
  \includegraphics[width=\linewidth]{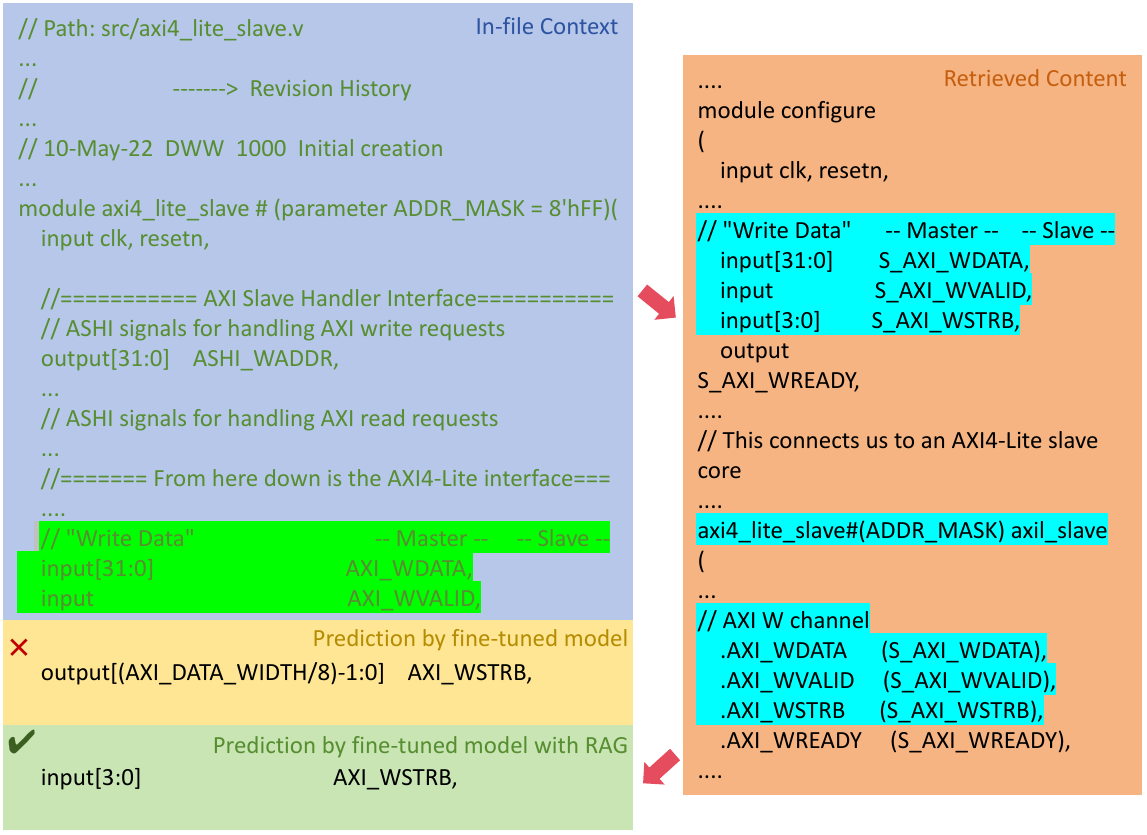}
  \caption{An example showing the prediction change without and with RAG. After retrieving the relevant code snippets, the model successfully outputs the correct code.}
  \label{fig:rag_code}
  
\end{figure}

To visually emphasize the importance of retrieval augmentation, we showcase an example of  an AXI4-Lite slave interface\cite{emu} in Figure~\ref{fig:rag_code}. Although the LLM predicts the code after fine-tuning, it generates incorrect input-output types and bit-widths. However, with the help of retrieval augmentation, the correct result is successfully predicted. Specifically, the embedding model retrieves the relevant content from the repository (highlighted in orange), which contains the instantiation of the incomplete 'axi4\_lite\_slave' module. According to the blue-highlighted content in the retrieval content, the 'AXI\_WSTRB' port is connected to 'S\_AXI\_WSTRB', which allows the correct result to be inferred from the definition of 'S\_AXI\_WSTRB'. As seen from the example above, the retrieved relevant code snippets effectively assist the LLM in predicting the correct RTL code.

\subsubsection{Embedding Model}
To validate the effectiveness of introducing long-context embedding model, we conduct experiments on samples exceeding 10k tokens, using different embedding models and chunk sizes. As shown in Table~\ref{tab:embedding_model}, the embedding model bge-large-en-v1.5\protect\cite{xiao2024c}, which only supports a context size of 512 tokens, results in significantly weaker performance compared to the embedding model jina-embeddings-v2-base-en\protect\cite{gunther2023jina} that supports a context size of 8192 tokens. Although performance improves with a larger chunk size when using bge-large-en-v1.5\protect\cite{xiao2024c}, the chunks cannot be effectively encoded due to the embedding model's context limitations. This results in inaccurate retrieval, which in turn undermines the performance of the LLM. In conclusion, it is essential to incorporate embedding models that support long contexts for repository-level RTL code completion task, which requires longer chunks and accurate retrieval.

\begin{table}[ht]

    \centering
    \begin{tabular}{lcccc}
        \hline
        \textbf{Embedding Model} & \textbf{Chunk Size} & \textbf{ES} & \textbf{EM} \\
        \hline
        \multirow{2}{*}{Bge-large-en-v1.5} & 512 & 80.6 & 48.4 \\
        & 4096 & 80.9 & 51.0 \\
        \hline
        \multirow{2}{*}{Jina-embeddings-v2-base-en} & 512 & 80.9 & 49.2 \\
        & 4096 & \textbf{82.8} & \textbf{53.1} \\
        \hline
    \end{tabular}
    \caption{Performance on samples exceeding 10k tokens of context with different embedding model and chunk sizes. Bge-large-en-v1.5\protect\cite{xiao2024c}, which supports 512 token length, is employed as a short context retrieval model, while the jina-embeddings-v2-base-en\protect\cite{gunther2023jina}, which supports 8192 token length, is employed as a long context retrieval model. The results are presented as percentages(\%), with the top performance emphasized in bold.}
    \label{tab:embedding_model}
\end{table}


	
		


\subsubsection{Splitting Strategy}
\begin{figure}[htbp]
  \centering
  \includegraphics[width=.4\linewidth]{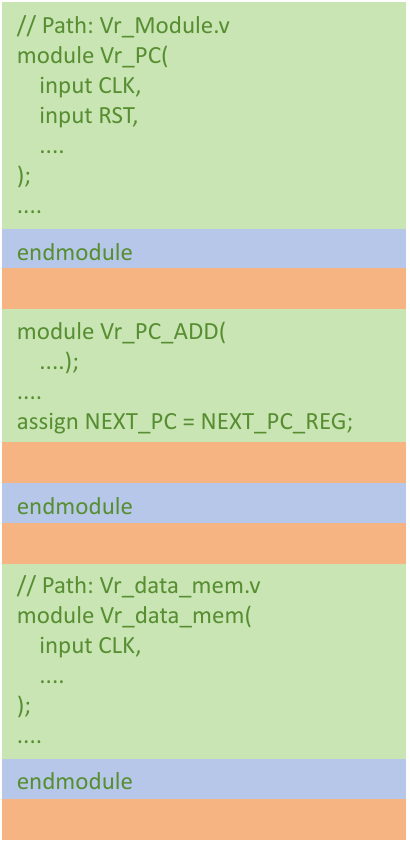}
  \caption{An illustration showcasing the effects of using different splitting keywords. Blue, green, and orange lines correspond to the dividing points of the keywords 'endmodule', '\textbackslash n', and '\textbackslash n\textbackslash n', respectively. To enhance visualization, we overlay green when other colors appear.}
  \label{fig:split}
  
\end{figure}

In the splitting process of RAG, the code within the repository $C_{repo}$ is initially partitioned into smaller pieces based on keywords. These pieces are then merged into chunks according to the predefined chunk size. Based on the characteristics of Verilog, we test three candidate split keywords: 'endmodule', '\textbackslash n' (line break), and '\textbackslash n\textbackslash n'. As shown in Figure~\ref{fig:split}, the '\textbackslash n' keyword corresponds to line-level splitting, which is very fine-grained and flexible. The 'endmodule' keyword, on the other hand, corresponds to segmentation at the module level, helping maintain the integrity of modules. The '\textbackslash n\textbackslash n' keyword falls between the above two, dividing content  within modules, between modules, and across files.

\begin{table}[ht]
    \centering
    \begin{tabular}{ccccc}
        \hline
        \textbf{Splitting Keyword} & \textbf{ES} & \textbf{EM} \\
        \hline
        \textbackslash n \textbackslash n & 82.1 & 52.9 \\

        endmodule & 82.7 & 54.4 \\
        \textbackslash n & \textbf{83.5} & \textbf{54.9} \\
        \hline
    \end{tabular}
    \caption{Performance on samples exceeding 10k tokens of context with different splitting keywords. The results are presented as percentages(\%), with the top performance emphasized in bold.}
    \label{tab:split}
\end{table}

Table~\ref{tab:split} shows the experimental results. Clearly, the '\textbackslash n \textbackslash n' keyword performs the worst, likely due to its uneven splits that disrupt the code internal  structure. The performance difference between 'endmodule' and '\textbackslash n' segmentation is not very significant. Both segmentation methods have their own advantages: 'endmodule' preserves the integrity of individual modules, while '\textbackslash n' allows for more uniform and flexible splitting, helping to avoid retrieving excessive irrelevant content.

\subsubsection{Chunk Size}

\begin{figure}[!h]
  \centering
  \includegraphics[width=.8\linewidth]{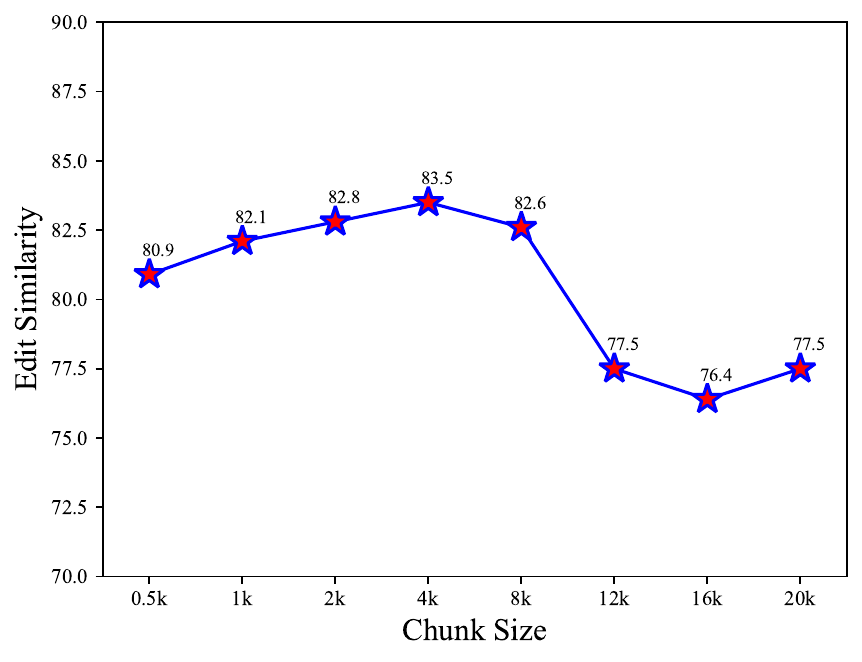}
  \caption{Edit Similarity of RAG system with different chunk sizes on samples exceeding 10k tokens of context. The results are presented as percentages(\%).}
  \label{fig:chunk}
  
\end{figure}

We conduct experiments to investigate the impact of chunk size on repository-level RTL code completion tasks. As shown in Figure~\ref{fig:chunk}, performance generally increases at first and then declines as the chunk size grows. This phenomenon is consistent with common understanding: Overly dense splitting disrupts the contextual information of the code, resulting in a deterioration of subsequent retrieval. On the other hand,  excessively large chunks introduce redundant information, which adversely affects the performance. What's more, our experiment shows that performance remains relatively stable when the chunk size is between 1k and 8k, confirming the robustness of our approach.

\subsection{Ablation Study of the Initial Decision-making Process}

As shown in Figure~\ref{fig:pipeline}, the decision on whether to use RAG is made at the beginning of the overall pipeline, based on whether the context length exceeds the predefined length. As a result, we can avoid the unnecessary overhead of applying RAG to every samples. Here we conduct an experiment to explore the impact of this initial decision-making phase. Based on the results in Table~\ref{tab:decision}, whether the initial decision phase is included or not does not significantly affect performance. In fact, using RAG on samples within the context length threshold almost only changes the internal arrangement of the code, which, according to the experimental results, does not significantly affect the LLM's performance on this task. Therefore, by applying RAG only to samples that exceed the context length threshold, we can reduce costs without compromising accuracy. 

\begin{table}[ht]
    \centering
    \begin{tabular}{ccccc}
        \hline
        \textbf{Initial Decision-making} & \textbf{ES} & \textbf{EM} \\
        \hline
        $\times$ & 84.2 & 55.8 \\

        \checkmark & 84.3 & 55.8 \\
        \hline
    \end{tabular}
    \caption{Performance with and without initial decision-making process. Not making an initial decision means applying RAG to all samples without distinction. The results are presented as percentages(\%).}
    \label{tab:decision}
\end{table}

\section{Conclusion}
In this article, we present RTLRepoCoder, the first solution specifically designed for repository-level RTL code completion, aimed at advancing the application of LLMs in real-world hardware design. First, we fine-tune an LLM using open-source Verilog repositories on GitHub with an extended context size, optimizing it for long-context Verilog code generation. Next, for samples exceeding the predefined context size, we employ an embedding model to retrieve relevant code snippets, so as to enhance the LLM's prediction accuracy.  The fine-tuning process, which boosts the LLM's capacity within its context window, works in synergy with the retrieval augmentation that increases the relevant information available within the input. By combining these two approaches, our solution achieves SOTA performance on public evaluation benchmark, significantly surpassing GPT-4 and advanced domain-specific LLMs in Edit Similarity and Exact Match metrics. Comprehensive quantitative experiments and qualitative analyses demonstrate the effectiveness of our design. We hope that our work can assist real-world hardware design and  pave the way for deeper integration between LLMs and the field of hardware design in the future.

\appendix





\bibliographystyle{named}
\bibliography{ijcai25}

\end{document}